\documentclass[aps,showpacs,twocolumn,preprintnumbers,amsmath,amssymb]{revtex4}
\usepackage{graphicx}
\usepackage{bm}

\def\br{ \bm{r} }
\def\bk{ \bm{k} }
\def\bo{ \bm{0} }
\def\im{ \,\mathrm{Im}\,}
\def\re{ \,\mathrm{Re}\,}
\def\Tr{ \,\mathrm{Tr}\,}

\begin{document}

\title{Effect of disorder on the NMR relaxation rate 
in two-band superconductors}

\author{ B. Mitrovi\'c and K. V. Samokhin}

\affiliation{Department of Physics, Brock University,
St.Catharines, Ontario, Canada L2S 3A1}
\date{\today}

\begin{abstract}
We calculate the effect of nonmagnetic impurity scattering on
the spin-lattice relaxation rate in two-band superconductors with 
the {\it s}-wave pairing symmetry. It is found that for the interaction 
parameters appropriate for MgB$_2$ the Hebel-Slichter peak is 
suppressed by disorder in the limit of small interband impurity 
scattering rate. In the limit of strong impurity scattering,
when the gap functions in the two bands become nearly equal,
the single-band behavior is recovered with a well-defined coherence 
peak just below the transition temperature.
\end{abstract}


\maketitle

\section{Introduction}
\label{sec: Introduction}

The discovery of superconductivity in MgB$_2$ \cite{MgB2} has 
revived the interest in two-band superconductors \cite{SMW59,Moskal59}. 
Most experiments on this compound \cite{MgB2-exp-review} are 
consistent with two distinct superconducting gaps 
$\Delta_\pi$ and $\Delta_\sigma$, with 
$\Delta_\sigma/\Delta_\pi\simeq 2.63$ \cite{Golub02}. The only
exception seems to be the nuclear magnetic resonance (NMR) 
relaxation rate $T_1^{-1}$ \cite{Kote01,Kote02} in the superconducting state. 
The NMR relaxation rate 
measurements on ${}^{11}$B nucleus by Kotegawa {\em et al.} 
\cite{Kote01,Kote02} found a small Hebel-Slichter coherence peak 
\cite{HS59} just below the superconducting transition temperature
$T_{c}$, although there are some questions as to 
whether the peak  is present at all \cite{Jung01,Baek02}. Moreover, 
by assuming that the contributions to the relaxation rate from the 
two bands are simply additive Kotegawa {\em et al.} found \cite{Kote02}
a better fit to their data on ${}^{11}$B nuclei by assuming that only 
one band participates in the nuclear relaxation process.

In a recent work \cite{SM05} we have examined the NMR relaxation rate in 
both conventional and unconventional {\em clean} two-band superconductors.
The essence of our findings is that if the Fermi contact interaction 
between the nucleus and the conduction electrons dominates the 
relaxation mechanism, then
the measurements of $T_1^{-1}$ probe the
properties of the electron subsystem which are local in real space
and hence extremely non-local in the momentum space
\cite{Slichter90}. We showed that for a two-band superconductor
this gives rise to inter-band interference terms in
$T_1^{-1}$, in addition to direct contributions from each band.  

Here we consider the effect of nonmagnetic impurity scattering on 
$T_1^{-1}$ for a two-band superconductor. It is well known \cite{Golub_Mazin,
Mitro04-2,Nicol05,Dolgov05} that the interband scattering by nonmagnetic impurities 
reduces the transition temperature of an {\it s}-wave two-band superconductor. 
In the case of MgB$_2$ it has been argued \cite{Erwin_Mazin} that one 
requires defects which produce lattice distortions in order to achieve 
large enough interband scattering which would lead to a significant 
suppression of $T_{c}$. Experimentally, such reduction of $T_{c}$ was 
observed on samples of MgB$_2$ irradiated by fast neutrons \cite{Wang,Putti06,DiCapua06,
Daghero06}.
However, there is a finite interband scattering by impurities 
even in unirradiated and undoped MgB$_2$ as evidenced by 
the break-junction tunneling experiments \cite{SZGH} whose interpretation 
was essentially based on the work of Schopohl and Scharnberg \cite{SS77} 
on the tunneling density of states of a disordered two-band superconductor.
The interband scattering rate from the low gap band to the high gap band 
used to fit the tunneling data in \cite{SZGH} was comparable to the transition 
temperature of 39 K. Thus it seems reasonable to investigate the effect of disorder 
on $T_1^{-1}$ over a wide range of scattering rates.

The NMR relaxation rate in a superconductor is 
determined by both normal and anomalous local density of states. In the 
case of a two-band superconductor these quantities depend on the interband 
scattering \cite{SS77,Golub_Mazin,Dolgov05} and on temperature \cite{Dolgov05} 
in a rather complex way and one cannot a priori guess the effect of disorder 
on the temperature dependence of $T_1^{-1}$ except in the limits of a small 
and very large interband impurity scattering rate. In the former case the 
peaks in the partial densities of states are somewhat reduced and broadened 
by the interband scattering compared to the clean system and one would expect the 
Hebel-Slichter coherence peak in $T_1^{-1}$ below $T_{c}$ to be reduced and 
broadened by disorder. In the limit of a very large interband scattering such 
that the gap functions in the two bands become nearly equal  
(the Anderson limit \cite{Anderson59}),
the difference between the local densities of states in the two bands has 
no consequences and 
one expects to find the temperature dependence of $T_1^{-1}$ characteristic of 
a single-band superconductor, with the size of the Hebel-Slichter peak determined 
by the strength of the electron-phonon coupling in the system \cite{AR91,AC91}. 
In the case of MgB$_2$, which is a medium coupling superconductor \cite{Golub02}, 
the Hebel-Slichter peak is expected to be quite large in the Anderson limit.   
Indeed we find that our detailed numerical calculations of $T_1^{-1}$ over 
a wide range of the interband impurity scattering rates confirm such a trend.

The rest of the article is organized as follows. In
Sec. \ref{sec: Theory}, we summarize the strong coupling theory 
of the NMR relaxation rate in a disordered {\it s}-wave two-band 
superconductor assuming that the Fermi contact interaction between 
the nuclear spin and the conduction electrons provides the dominant 
relaxation mechanism. In Sec. \ref{sec: Results} we give the 
results of our numerical calculations using the interaction parameters 
for MgB$_2$ \cite{Golub02,Mitro04-2,Mitro04-1} and in Sec. \ref{sec: 
Conclusions} we give conclusions.

\section{Theory}
\label{sec: Theory}

The relaxation rate of a nuclear spin 
due to the hyperfine
contact interaction with the band electrons is given by \cite{SM05}
\begin{equation}
\label{T1T general}
    R\equiv\frac{1}{T_1T}=-\frac{J^2}{2\pi}\lim\limits_{\omega_0\to
    0}\frac{\im K^R_{+-}(\omega_0)}{\omega_0},
\end{equation}
where $J$ is the hyperfine coupling constant, $\omega_0$ is the
NMR frequency, and $K^R_{+-}(\omega_0)$ is the analytic continuation 
of the Fourier transform $K(\nu_m)$ of the imaginary time correlator
\begin{equation}
\label{K_imaginary time}
K(\tau)=-\langle\langle T_{\tau}(S_+(\bo,-i\tau)S_-(\bo,0))\rangle\rangle_{i
}\>
\end{equation}
averaged over the impurity configurations.
Here $S_\pm(\br,-i\tau)=e^{H_e\tau}S_\pm(\br)e^{-H_e\tau}$, $H_e$ is the
electron Hamiltonian, and
\begin{equation}
\label{Spm}
    S_+(\br)=\psi^\dagger_\uparrow(\br)\psi_\downarrow(\br),\quad
    S_-(\br)=\psi^\dagger_\downarrow(\br)\psi_\uparrow(\br)
\end{equation}
with $\psi^\dagger_\sigma(\br)$ and $\psi^\dagger_\sigma(\br)$ being the  
electron field operators
($\hbar=k_B=1$ in our units, and we consider a system of unit volume). 
$H_e$ contains both electron-phonon and screened Coulomb interactions as 
well as the scattering off randomly located nonmagnetic impurities.

We assume that there are two spin-degenerate electron bands in the
crystal (the generalization to an arbitrary number of bands is
straightforward), and neglect the spin-orbit coupling. Then
the spin operators
(\ref{Spm}) can be written in the band representation, using
\begin{equation}
\label{psis}
\psi_\alpha(\br)=\sum\limits_{i,\bk}e^{i\bk\br}u_{i,\bk}(\br)c_{i,\bk\alpha}\>,
\end{equation}
where $u_{i,\bk}(\br)$ are the Bloch functions, which are periodic
in the unit cell. Inserting these
into Eqs. (\ref{Spm}), one obtains 
\begin{eqnarray}
\label{strong_Kpm nu m}
 K(\nu_m) & = &\frac{1}{2}T\sum_{n}\sum_{\bk_{1}\bk_{2}}\sum_{i,j}
	   |u_{i,\bk_1}(\bo)|^2|u_{j,\bk_2}(\bo)|^2 
	   \nonumber \\
     & &  \times \Tr[\hat
G_i(\bk_1,\omega_n)i\tau_2\hat G_j(\bk_2,-(\omega_n+\nu_m)) 
 \nonumber \\
     & &  \times \hat{\Gamma}_{ij}(\bk_1,\bk_2;\omega_n,\nu_m)]\>,
\end{eqnarray}
where $\hat G_i(\bk,\omega_n)$ are the impurity-averaged Green's functions
given by
\begin{equation}
\label{strong_Nambu GFs omega}
    \hat G_i(\bk,\omega_n)=-\frac{i\omega_n
    Z_{i}(\omega_n)\tau_0+\xi_{i,\bk}\tau_3+\phi_{i}(\omega_n)\tau_1}
    {\omega_n^2Z^2_{i}(\omega_n)+\xi_{i,\bk}^2+\phi_{i}^2(\omega_n)}\>.
\end{equation}
Here $Z_{i}(\omega_n)$
and $\phi_{i}(\omega_n)$ are the renormalization function and
the pairing self-energy, respectively, for the $i$th band.

In Ref.~\cite{SM05} we argued that the contribution to the 
vertex functions
$\hat{\Gamma}_{ij}(\bk_1,\bk_2;\omega_n,\nu_m)$
from the electron-phonon interaction can be ignored while the 
effect of the Coulomb interaction drops out from the ratio 
$R_s/R_n$. In calculating the contibution to $\hat{\Gamma}_{ij}$ 
from the impurity scattering in the conserving approximation \cite{Schr}
one considers only the ladder impurity diagrams since the self-energies
are calculated in the self-consistent second Born approximation.
We have shown recently \cite{SM06} that in the case of a single 
three-dimensional band
the contribution to the vertex function $\hat{\Gamma}$ from 
the ladder impurity diagrams 
is of the order $\gamma/\epsilon_{F}$, where $\gamma$  
is the impurity scattering
rate and $\epsilon_{F}$ is the Fermi energy, and that 
the impurity vertex
corrections can therefore be neglected.
The same applies to the contribution to $\hat{\Gamma}_{ij}$
from the ladder impurity diagrams  
since the structure of the impurity 
ladder diagrams for the $\hat{\Gamma}_{ij}$s is completely analogous to 
the single-band case. 
Hence, we replace
$\hat{\Gamma}_{ij}$ in Eq. (\ref{strong_Kpm nu m}) with 
matrix $i\tau_2$ in computing the ratio of the spin-lattice
relaxation rates in the superconducting and normal states. 

Next, by introducing the spectral representation for $\hat
G_i(\bk,\omega_n)$
one can  calculate the Matsubara sums in Eq.
(\ref{strong_Kpm nu m}) and then analytically continue the result 
to just above the real frequency axis $i\nu_{m}\to\omega_0+i0^+$ \cite{SM05}. 
In the limit $\omega_0\to 0$ we obtain
\begin{equation}
\label{strong_T1T final result}
    \frac{R_s}{R_n}=2\int\limits_0^{+\infty}d\omega
    \left(-\frac{\partial f}{\partial\omega}\right)
    \frac{N^2(\omega)+M^2(\omega)}{N_n^2}\>,
\end{equation}
where the densities of states (both normal and anomalous) are defined by 
\begin{eqnarray}
\label{Ns ab strong}
    N(\omega)&=&\sum_iN_{n,i}\re\frac{\omega}{\sqrt{\omega^2-\Delta_i^2(\omega)}}\>, \\
\label{Ms ab strong}
    M(\omega)&=&\sum_iN_{n,i}\re\frac{\Delta_i(\omega)}{\sqrt{\omega^2-
    \Delta_i^2(\omega)}}\>.
\end{eqnarray}
Here
\begin{equation}
\label{Nn ab}
    N_{n,i}=N_{F,i}\left\langle|u_{i,\bk}(\bo)|^2\right\rangle_i
\end{equation}
are the local densities of states at the nuclear site in the 
normal state, 
with $N_{F,i}$ the Fermi level density of states in $i$th band, 
$N_n=N_{n,1}+N_{n,2}$,  
and $\Delta_i(\omega)=\phi_i(\omega)/Z_i(\omega)$ is the gap
function in $i$th band. 

The gap functions are obtained by solving the finite temperature 
Eliashberg equations on the real axis \cite{Scalapino69}, 
which include the electron-phonon interaction, screened Coulomb 
repulsion and both the interband and intraband impurity scattering 
described by the self-consistent second Born approximation:
\begin{eqnarray}
\label{Eli1}
\phi_{i}(\omega) & = &  \phi_{i}^0(\omega)+i\sum_{j}\frac{\gamma_{ij}}{2}
\frac{\Delta_j(\omega)}{\sqrt{{\omega}^2-\Delta_j^2(\omega)}}\>,
                                             \\
\label{Eli2}
\phi_{i}^0(\omega) &  = &  \sum_j\int\limits_0^{\omega_c}d\omega'
\re\frac{\Delta_j(\omega')}{\sqrt{{\omega'}^2-\Delta_j^2(\omega')}}
                      \nonumber \\
& &
\times\left[f(-\omega')K_{+,ij}(\omega,\omega')-f(\omega')
\right.
          \nonumber \\
& &
\times K_{+,ij}(\omega,-\omega')
-\mu_{ij}^{*}(\omega_c)\tanh\frac{\omega'}{2T} \nonumber \\
 & & \left. +K_{+,ij}^{TP}(\omega,\omega')-K_{+,ij}^{TP}(\omega,-\omega')
\right]\>,  \\
\label{Eli3}
Z_{i}(\omega) & = & Z_{i}^0(\omega)+i\sum_{j}\frac{\gamma_{ij}}{2}
\frac{1}{\sqrt{{\omega}^2-\Delta_j^2(\omega)}}\>,  \\
\label{Eli4}
Z_{i}^0(\omega) & = & 1-\frac{1}{\omega} \sum_j\int\limits_0^{+\infty}d\omega'
\re\frac{\omega'}{\sqrt{{\omega'}^2-\Delta_j^2(\omega')}}
			    \nonumber \\
& & \times
\left[f(-\omega')K_{-,ij}(\omega,\omega')-f(\omega')K_{-,ij}(\omega,-\omega')
\right. \nonumber \\
 & & \left. +K_{-,ij}^{TP}(\omega,\omega')+K_{-,ij}^{TP}(\omega,-\omega')
\right]\>,
\end{eqnarray}
where
\begin{eqnarray}
\label{kernel}
    K_{\pm,ij}(\omega,\omega')   =
   \int\limits_0^{+\infty}d\Omega\;\alpha^{2}F_{ij}(\Omega) \nonumber \\
  \times\left[\frac{1}{\omega'+\omega+\Omega+i0^{+}}
 \pm  \frac{1}{\omega'-\omega+\Omega-i0^{+}}\right]\>, \\
\label{TPkernel}
    K_{\pm,ij}^{TP}(\omega,\omega')  =
   \int\limits_0^{+\infty}d\Omega\;
   \frac{\alpha^{2}F_{ij}(\Omega)}{e^{\Omega/T}-1} \nonumber \\
   \times\left[\frac{1}{\omega'+\omega+\Omega+i0^{+}}
\pm   \frac{1}{\omega'-\omega+\Omega-i0^{+}}\right]\>.
\end{eqnarray}
Here $\alpha^{2}F_{ij}(\Omega)$ and $\mu_{ij}^{*}(\omega_c)$ are 
intraband and interband electron-phonon coupling functions and 
Coulomb 
\begin{figure}[t]
\includegraphics[angle=0,width=8cm]{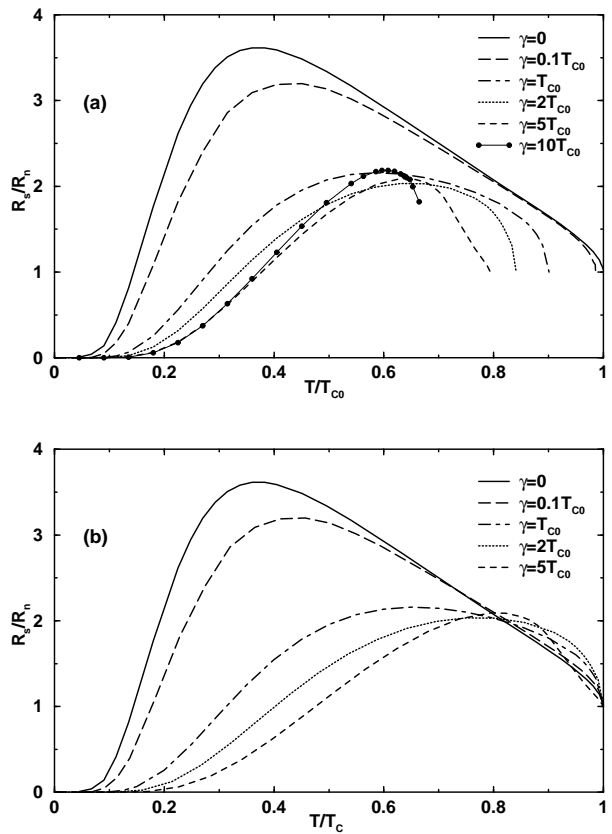}
\caption{The ratio $R_{s}/R_{n}$ in the case when the
relaxation is dominated by the lower-gap band ($N_{n,\sigma}$=0) 
for several values of the interband scattering rate 
$\gamma_{\pi\sigma}\equiv\gamma$ in the units of
transition temperature of the clean system $T_{c0}$. In (a) 
$R_{s}/R_{n}$ is plotted as a function of temperature
in units of $T_{c0}$. Note that for $\gamma/T_{c0}$=10
the results do not extend all the way to the transition 
temperature $T_{c}$ because of the poor convergence of 
the real-axis Eliashberg equations near $T_{c}$ for such 
a high value of $\gamma$. 
In (b)
$R_{s}/R_{n}$ is plotted as a function of $T/T_{c}$, where 
$T_{c}$ is the transition temperature of disordered system, 
in order to illustrate the change in the relative width 
and position of the coherence peaks with increasing $\gamma$.
} \label{fig:fig1}
\end{figure}
repulsion parameters for the cutoff $\omega_{c}$, respectively.
The impurity scattering rates are defined by $\gamma_{ij}/2=
n_{imp}\pi N_{F,j}(0)|V_{ij}|^2$ where $n_{imp}$ is the concentration of 
impurities 
and $V_{ij}$ is the Fermi surface averaged 
matrix element of the change in the lattice
potential caused by an impurity between the states 
in the bands $i$ and $j$.

It is easy to see that the intraband scattering rates $\gamma_{ii}$ 
drop out from the equations for the gap functions
$\Delta_i(\omega)=\phi_i(\omega)/Z_i(\omega)$ which are obtained 
from Eqs.~(\ref{Eli1}-\ref{Eli4}), and only the interband impurity 
scattering affects the gap functions and hence $R_s/R_n$, 
Eqs.~(\ref{strong_T1T final result}-\ref{Ms ab strong}).
We should point out that the Eqs.~(\ref{Eli1}-\ref{Eli4}) are the 
same as the strong coupling equations for the McMillan tunneling 
model of the superconducting proximity effect \cite{McMillan68}
and were first solved numerically at zero temperature by Zarate 
and Carbotte \cite{ZC85} over twenty years ago.

\section{Results}
\label{sec: Results}

In order to examine the effect of the interband impurity scattering 
on the NMR relaxation rate of a singlet two-band superconductor
we have calculated $R_s/R_n$ for the interaction parameters of 
MgB$_{2}$. The four electron-phonon coupling functions 
$\alpha^{2}F_{ij}(\Omega)$, $i,j=\sigma,\pi$, were calculated by 
Golubov {\em et al.} \cite{Golub02} and the Coulomb 
repulsion parameters $\mu_{ij}^*(\omega_c)$ were determined in 
\cite{Mitro04-2,Mitro04-1} based on the screened 
Coulomb repulsion parameters of 
MgB$_{2}$ calculated by 
Choi {\em et al.} \cite{Choi04}. Since $\gamma_{\sigma\pi}/
\gamma_{\pi\sigma}=N_{F,\pi}/N_{F,\sigma}$ there is only 
one independent interband scattering rate parameter and we 
choose $\gamma_{\pi\sigma}\equiv\gamma$.
\begin{figure}[t]
\includegraphics[angle=0,width=8cm]{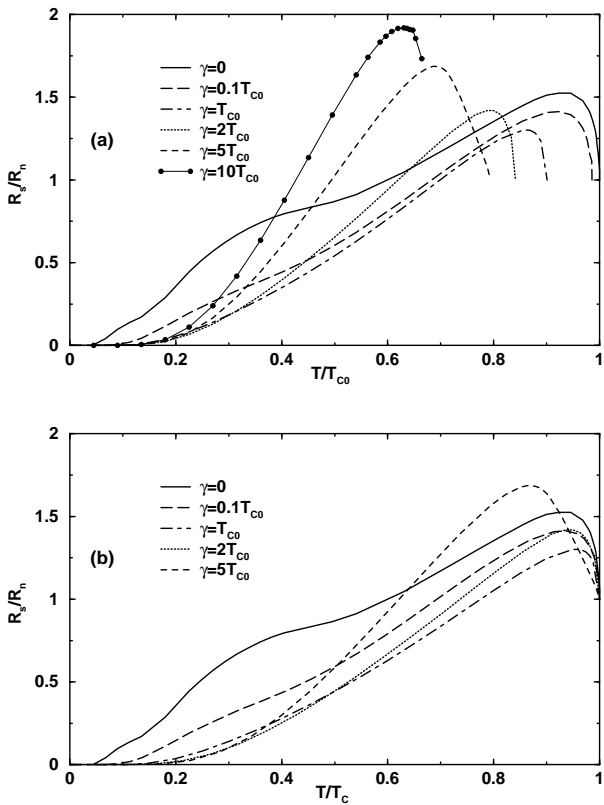}
\caption{The ratio $R_{s}/R_{n}$ 
for $N_{n,\pi}/N_{n,\sigma}$=0.45 for
several values of the interband scattering rate
$\gamma_{\pi\sigma}\equiv\gamma$ in the units of the 
transition temperature of the clean system $T_{c0}$. In (a)
$R_{s}/R_{n}$ is plotted as a function of temperature
in units of $T_{c0}$.
In (b)
$R_{s}/R_{n}$ is plotted as a function of $T/T_{c}$, where
$T_{c}$ is the transition temperature of disordered system. 
} \label{fig:fig2}
\end{figure}
Our representative results are shown in  
Figs.~\ref{fig:fig1} and \ref{fig:fig2}.
Figure \ref{fig:fig1} is our theoretical prediction for the relaxation 
rates on ${}^{25}$Mg nuclei in MgB$_2$, for which 
the dominant relaxation mechanism is the Fermi contact interaction 
\cite{BAR01,PM01} and the electronic structure calculations \cite{Bose05} 
give $N_{n,\sigma}\approx$0 for the Mg-site.
To the best of our knowledge there are no measurements of 1$/T_{1}$ 
in the superconducting state on 
${}^{25}$Mg in MgB$_2$, presumably because of its 
small magnetic moment and a low natural abundance, but the
experiments performed in \cite{Mali02,Gerash02} indicate that it
is possible to measure $^{25}R$ below the superconducting
transition temperature. Such measurements would be highly 
desirable since our theory is quantitatively correct for 
${}^{25}$Mg nucleus. The broad peaks in Fig.~\ref{fig:fig1} between  
0.3$T_{c}$ and 0.6$T_{c}$ for 0$\leq\gamma\leq$2$T_{c}$ are analogous to the 
broad peaks found in the microwave conductivity of MgB$_2$ \cite{Jin03} 
in the same temperature range. In \cite{Jin03} the best fits to the 
data were obtained by assuming that the $\pi$-band gives the dominant 
contribution to the microwave conductivity and our results in Fig.~\ref{fig:fig1} 
also give only the $\pi$-band contribution to $R_{s}/R_{n}$ 
($N_{n,\sigma}$ = 0). Since both NMR relaxation rate and the 
microwave conductivity have the same coherence factors in the single band 
case \cite{Tink96} the similarity between our prediction for $^{25}R$ and 
the results obtained in \cite{Jin03} is not accidental. 

In Fig.~\ref{fig:fig2} we present our results for 
$N_{n,\pi}/N_{n,\sigma}$ = 0.45, which would correspond to 
${}^{11}$B nucleus \cite{Bose05} in MgB$_2$ if the dominant relaxation 
mechanism were the Fermi contact interaction. However, 
the local-density approximation (LDA) calculations \cite{BAR01,PM01} 
have found that at the ${}^{11}$B 
nucleus the most significant contribution to the relaxation comes 
from the interaction with the electronic orbital part of the 
hyperfine field. Hence, our results in Fig.~\ref{fig:fig2} 
should not be compared directly with the experimental results 
for $^{11}R$. 
While the predictions in \cite{BAR01,PM01} were confirmed by experiments 
\cite{Mali02,Gerash02,Papa02} in the {\em normal} state, until 
recently \cite{KMS06} nothing was known theoretically about the temperature 
dependence of 1$/T_{1}$ in a superconductor in which the orbital part of the
hyperfine field dominates the NMR relaxation. Our preliminary results 
\cite{KMS06} for a single band superconductor indicate that in such a case 
the temperature dependence of $R_{s}/R_{n}$ is given by the standard 
expressions obtained for the Fermi contact interaction \cite{Scalapino69, 
Tink96} provided that $\gamma$ is much greater than $T_c$.  
In our treatment \cite{KMS06} of the orbital contribution to $R_{s}/R_{n}$ the 
extended nature of the single electron states  
participating in the formation of Cooper pairs
played the key
role, and it is not clear how the treatment of \cite{BAR01,PM01} 
in which a few localized orbitals are responsible for the orbital part of the
hyperfine field could be extended to the superconducting state. 
\begin{figure}[t]
\includegraphics[angle=0,width=8cm]{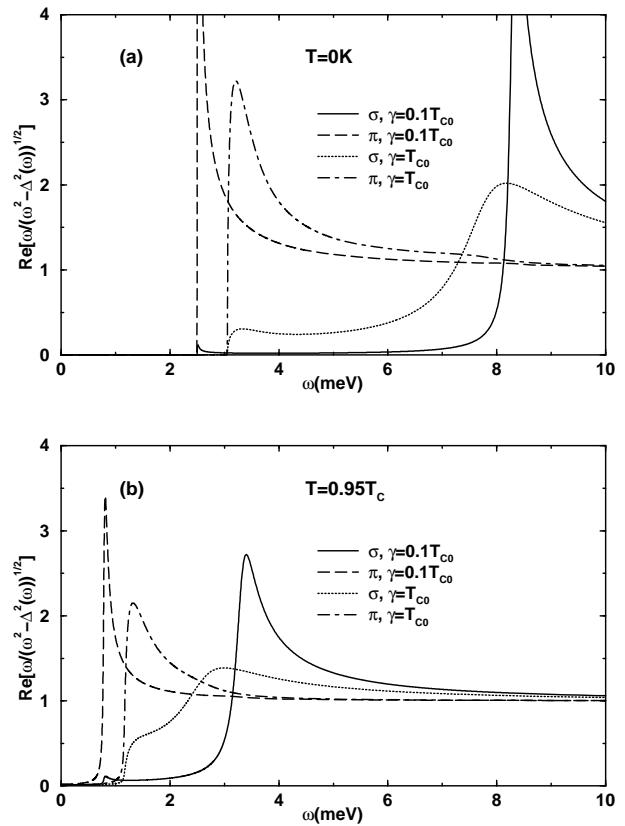}
\caption{The densities of states,
$\re[\omega/\sqrt{\omega^2-\Delta_{\sigma,\pi}(\omega)}]$,
in the $\sigma$ and $\pi$ bands at (a) $T$ = 0 and (b) $T$ = 0.95$T_{c}$, for the
impurity interband scattering rates $\gamma$ = 0.1$T_{c0}$ and
$\gamma$ = $T_{c0}$.
}
\label{fig:fig3}
\end{figure}

It is apparent from Figs.~\ref{fig:fig1} and \ref{fig:fig2} that 
as the interband impurity scattering rate 
increases initially from 0 to about 2$T_{c0}$ the 
coherence peak in the NMR relaxation rate is reduced and moved closer to 
the $T_{c}$. In the case of $^{25}R$ the peak is also broadened (see 
Fig.~\ref{fig:fig1}b) while the shoulder in $R_{s}/R_{n}$ for 
$N_{n,\pi}/N_{n,\sigma}$ = 0.45 in Fig.~\ref{fig:fig2} for $\gamma$=0 
resulting from the $\pi$-band contribution is rapidly reduced,  
Fig.~\ref{fig:fig1}. As we have anticipated in Sec. \ref{sec: Introduction} 
these changes in $R_{s}/R_{n}$ at low values of $\gamma$ are a direct consequence 
of the changes in both normal and anomalous densities 
of states in the superconducting state
with increasing impurity scattering,
Eqs.~(\ref{strong_T1T final result}-\ref{Ms ab strong}).
In 
Fig.~\ref{fig:fig3} we 
show the partial densities of states in the two bands at $T$ = 0 
and at $T$ just below the $T_{c}$, for two lowest values of 
the interband impurity scattering parameter $\gamma$ from 
Figs.~\ref{fig:fig1} and \ref{fig:fig2}. Clearly, in the low $\gamma$ 
regime increasing interband impurity scattering leads to a reduction 
and smearing of the partial
densities of states both at low and high temperatures. As a result the
the Hebel-Slichter coherence peaks are reduced in size and in the 
case of a single-band contribution, Fig.~\ref{fig:fig1}, the peak is 
also broadened. 
\begin{figure}[h]
\includegraphics[angle=0,width=8cm]{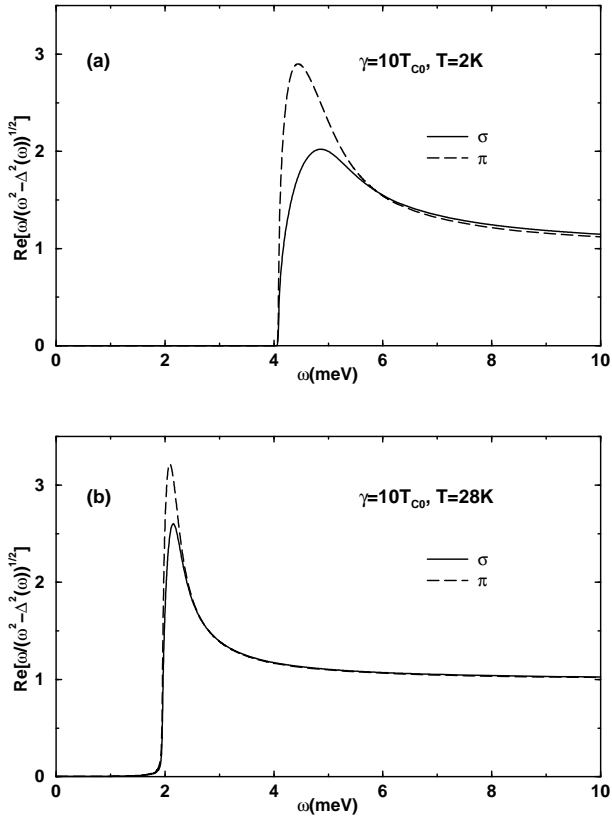}
\caption{The densities of states,
$\re[\omega/\sqrt{\omega^2-\Delta_{\sigma,\pi}(\omega)}]$, 
in the $\sigma$ and $\pi$ bands at (a) $T$ = 2 K and (b) $T$ = 28 K for the
highest impurity interband scattering rates $\gamma$ = 10$T_{c0}$
considered in this work.
}
\label{fig:fig4}
\end{figure}

At high vales of $\gamma$ (the Anderson limit) one expects that the gap 
functions in the two bands will become nearly equal, resulting in nearly 
equal normal and anomalous densities of states in both bands. As a result 
one expects the system to display $1/T_1$ characteristic of a single-band  
superconductor with pronounced Hebel-Slichter peak, barring extremely 
strong electron-phonon coupling \cite{AR91,AC91}. Indeed, as illustrated 
in Fig.~\ref{fig:fig4} for $\gamma$ = 10$T_{c0}$, the densities of 
states in the two bands are very similar, in particular at 
high temperature where the damping term in Eq.~(\ref{Eli1}) associated 
with the interband impurity scattering is reduced by the smaller gaps 
at high temperatures. The corresponding $R_{s}/R_{n}$ shown in 
Figs.~\ref{fig:fig1}(a) and \ref{fig:fig2}(a) have well pronounced 
coherence peaks similar in shape to what one would expect for a 
single-band {\it s}-wave superconductor.

\section{Conclusions}
\label{sec: Conclusions}

We calculated the NMR relaxation rate $1/T_1$ in an {\it s}-wave 
two-band superconductor with
impurities, assuming that the relaxation of nuclear spins is
controlled by the Fermi contact interaction with the band
electrons. Using the interaction parameters of MgB$_2$ \cite{Golub02,Choi04} 
we found that for low interband impurity
scattering rates disorder suppresses  
Hebel-Slichter coherence peak as a result of the smearing of 
the densities of states in the two bands with two different gaps.
For high scattering rates, as the gap functions in the two bands 
become nearly equal, the system behaves as a single-band superconductor 
with a well-developed coherence peak below $T_{c}$, as appropriate for 
a medium electron-phonon coupling parameter $\lambda\approx$1 and 
a small ratio $T_{c}/\Omega_{max}$, where $\Omega_{max}$ is the maximum 
phonon frequency. However, one should keep in mind that in the 
limit of strong disorder, such as found in the samples of MgB$_2$ irradiated 
with high neutron fluences \cite{Wang,Putti06,DiCapua06,
Daghero06},
our treatment of disorder is not sufficient to quantitatively describe 
all experimental results. Indeed recently observed \cite{Putti06,DiCapua06,Daghero06}
reduction of the smaller $\pi$-band gap with increasing disorder in 
neutron irradiated samples of MgB$_2$ can never be reproduced within the 
self-consistent second Born approximation treatment of impurity scattering 
used here and elsewhere in the literature on two-band superconductivity. 
For a highly disordered system
one would likely have to consider changes in the electron-phonon interaction and 
the normal state densities of states to account for the reduction of the gap 
in three-dimensional $\pi$-band.

\acknowledgments

We thank Anton Knigavko for helpful discussions and to S.~K.~Bose
for making available to us some of his electronic structure 
calculation results \cite{Bose05}.
This work
was supported by the Natural Sciences and Engineering Research
Council (NSERC) of Canada.


\begin{thebibliography}{99}

\bibitem{MgB2}
J. Nagamatsu, N. Nakagawa, T. Muranaka, Y. Zenitani, and J.
Akimitsu, Nature (London) \textbf{410}, 63 (2001).

\bibitem{SMW59}
H. Suhl, B. T. Matthias, and L. R. Walker, Phys. Rev. Lett.
\textbf{3}, 552 (1959).

\bibitem{Moskal59}
V. A. Moskalenko, Fiz. Met. Metalloved. \textbf{8}, 503 (1959).

\bibitem{MgB2-exp-review}
Review issue on MgB$_2$, edited by G. Crabtree, W. Kwok, P. C. Canfield, 
and S. L. Bud'ko, Physica C \textbf{385}, 1 (2003).

\bibitem{Golub02}
A. A. Golubov, J. Kortus, O. V. Dolgov, O. Jepsen, Y. Kong, O. K.
Andersen, B. J. Gibson, K. Ahn, and R. K. Kremer, J. Phys.:
Condens. Matter \textbf{14}, 1353 (2002).

\bibitem{Kote01}
H. Kotegawa, K. Ishida, Y. Kitaoka, T. Muranaka, and J. Akimitsu,
Phys. Rev. Lett. \textbf{87}, 127001 (2001).

\bibitem{Kote02}
H. Kotegawa, K. Ishida, Y. Kitaoka, T. Muranaka, N. Nakagawa, H.
Takagiwa, and J. Akimitsu, Phys. Rev. \textbf{66}, 064516 (2002).

\bibitem{HS59}
L. C. Hebel and C. P. Slichter, Phys. Rev. \textbf{113}, 1504
(1959).

\bibitem{Jung01}
J. K. Jung, S. H. Baek, F. Borsa, S. L. Bud'ko, G. Lapertot, and
P. C. Canfield, Phys. Rev. B \textbf{64}, 012514 (2001).

\bibitem{Baek02}
S. H. Baek, B. J. Suh, E. Pavarini, F. Borsa, R. G. Barnes, S. L.
Bud'ko, and P. C. Canfield, Phys. Rev. B \textbf{66}, 104510
(2002).

\bibitem{SM05}
K. V. Samokhin and B. Mitrovi\' c, Phys. Rev. B \textbf{72}, 134511 (2005).

\bibitem{Slichter90}
C. P. Slichter, \emph{Principles of Magnetic Resonance}, 3rd ed.
(Springer-Verlag, Berlin, 1990).

\bibitem{Golub_Mazin}
A. A. Golubov and I. I. Mazin, Phys. Rev. B \textbf{55}, 15146 (1997).

\bibitem{Mitro04-2}
B. Mitrovi\' c, J. Phys.: Condens. Matter \textbf{16}, 9031
(2004).

\bibitem{Nicol05}
E. J. Nicol and J. P. Carbotte, Phys. Rev. B \textbf{71}, 054501(2005).

\bibitem{Dolgov05}
O. V. Dolgov, R. K. Kremer, J. Kortus, A. A. Golubov and S. V. Shulga, Phys. Rev. B 
\textbf{72}, 024504 (2005).

\bibitem{Erwin_Mazin}
S. C. Erwin and I. I. Mazin, Phys. Rev. B \textbf{68}, 132505 (2003).

\bibitem{Wang}
Y. Wang, F. Bouquet, I. Sheikin, P. Toulemonde, B. Revaz, M. Eisterer,
H. W. Weber, J. Hinderer and A. Junod, J. Phys.: Condens. Matter \textbf{15},
883 (2003).

\bibitem{Putti06}
M. Putti, M. Affronte, C. Ferdeghini, P. Manfrinetti, C. Tarantini and 
E. Lehmann, Phys. Rev. Lett. \textbf{96}, 077003 (2006).

\bibitem{DiCapua06}
R. Di Capua, H. U. Aebersold, C. Ferdeghini, V. Ferrando, P. Orgiani, M. Putti,
M. Salluzzo, R. Vaglio and X. X. Xi, cond-mat/0606606.

\bibitem{Daghero06}
D. Daghero, A. Calzolari, G. A. Ummarino, M. Tortello, R. S. Gonnelli, 
V. Stepanov, C. Tarantini, P. Manfrinetti and E. Lehmann, cond-mat/0607515. 

\bibitem{SZGH}
H. Schmidt, J. F. Zasadzinski, K. E. Gray and D. G. Hinks D  Phys. Rev. Lett.
\textbf{88}, 127002 (2002); Physica C \textbf{385}, 221 (2003).

\bibitem{SS77}
N. Schopohl and K. Scharnberg, Solid State Commun. \textbf{22},
371 (1977).

\bibitem{Anderson59}
P. W. Anderson, J. Phys. Chem. Solids \textbf{11}, 26 (1959).

\bibitem{AR91}
P. B. Allen and D. Rainer, Nature \textbf{349}, 396 (1991).

\bibitem{AC91}
R. Akis and J. P. Carbotte, Solid State Comm. \textbf{78}, 393
(1991).

\bibitem{Mitro04-1}
B. Mitrovi\' c,  Eur. Phys. J. B \textbf{38}, 451 (2004).

\bibitem{Schr}
J. R. Schrieffer, \emph{Theory of Superconductivity} (W. A.
Benjamin, New York, 1964).

\bibitem{SM06}K. V. Samokhin and B. Mitrovi\' c, unpublished.

\bibitem{Scalapino69}
D. J. Scalapino in {\em Superconductivity}, ed. by R. D. Parks
(Marcel Dekker, New York, 1969), Vol. 1, pp. 466-501.

\bibitem{McMillan68}
W. L. McMillan, Phys. Rev. \textbf{175}, 537 (1968).

\bibitem{ZC85}
H. G. Zarate and J. P. Carbotte, J. Low Temp. Phys \textbf{59}, 19 (1985).

\bibitem{Choi04}
H. J. Choi, D. Roundy, H. Sun, M. L. Cohen, and S. G. Louie, 
Phys. Rev. B \textbf{69}, 056502 (2004).

\bibitem{BAR01}
K. D. Belaschenko, V. P. Antropov, and S. N. Rashkeev, Phys. Rev.
B \textbf{64}, 132506 (2001).

\bibitem{PM01}
E. Pavarini and I. I. Mazin, Phys. Rev. B \textbf{64}, 140504(R)
(2001).

\bibitem{Bose05}
S. K. Bose, unpublished (2005).

\bibitem{Mali02}
M. Mali, J. Roos, A. Shengelaya, H. Keller, and K. Conder, Phys.
Rev. B \textbf{65}, 100518 (2002).

\bibitem{Gerash02}
A. P. Gerashenko, K. N. Mikhalev, S. V. Verkhovskii, A. E. Karkin,
and B. N. Goshchitskii, Phys. Rev. B \textbf{65}, 132506 (2002).

\bibitem{Jin03}
B. B. Jin, T. Dahm, A. I. Gubin, Eun-Mi Choi, Hyun Jung Kim, Sung-IK Lee, 
W. N. Kang and N. Klein, Phys. Rev. Lett. \textbf{91}, 127006 (2003).

\bibitem{Tink96}
M. Tinkham, {\em Introduction to Superconductivity} (McGraw-Hill,
New York, 1996).

\bibitem{Papa02}
G. Papavassiliou, M. Pissas, M. Karayani, M. Fardis, S. Koutandos,
and K. Prassides, Phys. Rev. B \textbf{66} 140514 (2002).

\bibitem{KMS06}
A. Knigavko, B. Mitrovi\' c and K. V. Samokhin, cond-mat/0605420.


\end{thebibliography}
\end{document}